# Position estimation based on UWB swarm optimization and comparison against traditional trilateration

Vinish Yogesh [1,2,*], Bert-Jan van Beijnum [2], Jaap H. Buurke [1,2], and Chris T. M. Baten [1,2]

[1] Roessingh Research and Development, Roessinghsbleekweg 33B, 7522 AH Enschede, The Netherlands;
[2] Department of Biomedical Signals and Systems, University of Twente, Drienerlolaan 5, 7522 NB Enschede, The Netherlands;

[*] Correspondence: v.yogesh@utwente.nl/ v.yogesh@rrd.nl

**Abstract:** Ultra-wideband (UWB) is a promising technology for indoor position estimation for various localization applications of object swarms, such as in 3D analysis of human movement with multiple on-body sensors or a swarm of drones in an indoor environment. However, most UWB-only position estimation methods are based on a star topology, where the position of a mobile node is estimated using distances from several fixed anchors. These approaches ignore the valuable inter-node distance estimates, possible in a fully-connected 'Swarm' topology, which could provide more redundancy in the set of available distance estimates used for the position estimation. This would improve the accuracy and consistency of the position estimates. Also, published studies do not analyze how input measurement errors affect the final position estimates, which makes it difficult to assess the reliability under varying conditions. Therefore, this study first proposes a UWB swarm optimization-based position estimation method that utilizes all available internode distances to enhance accuracy and compare against the traditional trilateration method that utilizes the star configuration. All validations were done with synthetic UWB data, to enable testing all input error situations. The comprehensive error sensitivity analysis was conducted to evaluate its robustness under varying noise conditions. The proposed method consistently outperformed trilateration, with position estimation error around 5.7 cm for realistic UWB distance input estimates, while for higher noise conditions, the proposed method had errors around 6 cm lower than the trilateration method, which had position estimation errors around 19 cm. This study demonstrates the general potential of the Swarm optimization-based method for position estimation as a more accurate and consistent alternative to traditional star-based trilateration methods.

## 1 Introduction

Ultrawide-band (UWB) has emerged as one of the promising technologies for indoor localization and position estimation applications. This is mainly due to their relatively low cost, low power consumption, large bandwidth, and robust performance [1]. Besides, the UWB-based positioning currently offers the highest accuracy and can achieve sub-decimeter accuracy for indoor positioning among other wireless communication modalities used for positioning, such as Wi-Fi, ZigBee, Bluetooth, and radio frequency identification [2,3]. However, in practical applications, achieving higher accuracies is only feasible in clear line-of-sight (LOS) conditions, such as free unobstructed spaces [4]. In typical indoor environments, the UWB positioning accuracy is often degraded by the presence of various obstacles such as furniture, human bodies, and structural elements. These obstructions create non-line-of-sight (NLOS) conditions and introduce additional errors on top of the baseline noise and bias observed under LOS conditions, leading to inaccurate position estimations [5,6].

To improve positioning accuracy, UWB is often combined with other sensors such as Magnetic Inertial Measurement Unit (MIMU) [5,7], LiDAR [8], or vision-based systems [9] through data fusion techniques, such as (Extended-) Kalman filters. These multimodal data fusion-based approaches have significantly improved position estimation performance. Among these sensor combinations, the integrated UWB and MIMU sensors have shown immense potential and have been increasingly explored in recent research [7,10,11]. This is attributed to the complementary error characteristics between MIMU and UWB. Specifically, position estimation algorithms that rely on MIMU sensor data usually suffer from integration drift, leading to errors that increase over time [12,13]. However, they are generally unaffected by NLOS conditions as observed in UWB. In contrast, UWB sensors provide position data that is free from drift by using absolute distance measurements, while they are disrupted in NLOS environments [10,14]. Therefore, this specific combination of the UWB/MIMU achieves a higher accuracy than each of them used independently and is now widely utilized for positioning/localization applications such as pedestrian navigation, autonomous robotics, and drones.

Though the UWB/MIMU combination has been successfully applied for indoor localization, pedestrian navigation, and robotic applications, they have not yet been widely adopted for the 3D analysis of human movement (3D AHM) and clinical applications that demand very high-accuracy position estimates of multiple moving nodes on the human body [15]. In these high-accuracy applications, the position estimation errors are expected to be around 1 cm or lower [15], while the accuracy currently achieved by these integrated UWB/MIMU sensor systems is still not sufficient for such high-accuracy applications, with reported position estimation errors in pedestrian navigation around 10 cm or higher [1,15,16]. Studies suggest that the MIMU component in the widely adopted UBW/MIMU fusion approach has already been extensively optimized to achieve the current state-of-the-art accuracies [17]. As a result, to further enhance accuracy in such data fusion frameworks, it is required to improve UWB positioning accuracy itself. Enhancing UWB performance can be addressed in two aspects: (1) improving distance measurement of the UWB (UWB ranging) and (2) refining position estimation algorithms based on UWB distance measurements. Within the 1$^{st}$ aspect improvement, several studies have attempted to improve UWB ranging accuracy through calibration techniques and error modeling approaches [17-22], reducing distance estimation errors to a residual bias of approximately 0.5 cm [17], and with a random error component of around 5 cm [18]. This study focuses on accuracy enhancement on the 2$^{nd}$ aspect through using more effective computation strategies by developing and validating advanced algorithms to estimate position from distance measurements in a fully-connected UWB node Swarm.

Most UWB position estimation algorithms rely on distances measured between three or more fixed UWB sensor nodes at known positions, usually referred to as anchors, and mobile UWB nodes with unknown positions, referred to as mobile nodes, to estimate their positions. This configuration follows a star topology where distances are measured only between the fixed anchors and each mobile node in the network. The existing position estimation algorithms can be generally classified as linear positioning algorithms (non-iterative) and non-linear positioning algorithms (iterative/recursive) [23]. Linear algorithms, such as trilateration [24] and least squares approaches [25], estimate position based on geometric relationships. Non-linear or iterative algorithms, on the other hand, depend on advanced mathematical frameworks, such as Taylor series approximations [26] or filtering techniques like an Extended Kalman Filter (EKF) [27] and Unscented Kalman Filter (UKF) [28], which incorporate motion models or data from additional sensors to refine position estimates. However, despite these advancements, most existing methods still struggle to provide the high accuracy required for applications such as clinical human movement analysis, where position estimation errors around 1 cm are sought [15]. Reported errors for only UWB-based positioning for most of the studies are around 10 cm [29-34], with the lowest achieved position estimation error being around 3 cm [34], achieved in a static LOS measurement scenario. These reported errors exceed the acceptable thresholds even in LOS conditions for applications demanding precise localization, making it necessary to explore alternative approaches to improve accuracy and robustness.

One of the key limitations of existing UWB positioning algorithms is their dependence on a star topology, where distances are measured only between anchors and one or more mobile tags in the network. This approach ignores the valuable internode distance information between all the mobile nodes connected in the network that could be exploited to enhance position estimates. Theoretically, a minimum of three anchors is required for 3D positioning with traditional star configuration [35], while a redundancy in star configuration can be achieved with having a higher number of anchors [36,37]. This could potentially aid in improving accuracy through optimization-based techniques, however, this configuration with increased anchors requires a more sophisticated infrastructure and expensive setup [37]. In addition, studies have shown that though the increased anchors could provide redundancy to help improve position estimation, they also introduce more errors and uncertainty, thereby causing instability [38,39].

Therefore, to address these limitations while utilizing the minimal number of anchors (3 anchors), a swarm-based topology where all available internode distances (which also includes distances between all mobile nodes) are utilized for position estimation, was explored in this study. This could also help minimize the dependency on a large number of anchors and could make it possible to have redundant data with a minimum of three anchors and two or more mobile nodes. Additionally, while many existing UWB positioning algorithms have been validated, most studies do not specify the input distance measurement errors used during evaluation, and no analysis of how input measurement errors affect final position estimates is conducted. Without this information, it is unclear under what conditions the reported accuracy can be reproduced, making it difficult to assess the method's performance in real-world scenarios. Therefore, it is important to explicitly evaluate how varying input error conditions in the distance estimates influence the resulting position estimation accuracy to ensure applicability in real-world scenarios.

To address these gaps, in this study, first, a non-linear optimization-based algorithm that utilizes the redundant internode distance measurements available in a swarm topology with a minimal anchor setup to improve localization accuracy is proposed, the 'swarm optimization algorithm'. The proposed method is validated with synthetically generated UWB distance data, created through controlled experiments where a high-precision optical motion capture system is used as ground truth for evaluating the estimated positions. To further validate the effectiveness of our approach, we compare its performance against the widely used trilateration method. Following this, an input error sensitivity analysis to evaluate the robustness of both methods under varying noise and bias conditions was conducted. The novelty of this work lies in introducing and evaluating a swarm-based optimization algorithm to identify its capability to enhance the accuracy and consistency of UWB positioning, as well as to present a comprehensive error sensitivity analysis providing valuable insights into how different input error conditions impact both our proposed method and trilateration.

## 2  Methods

### 2.1  UWB node swarm configuration

In this study the measurement system configuration (Figure 1) includes a total of $n$ UWB nodes, where three nodes act as anchors which are placed at fixed, known positions within the measurement region (orange nodes in the figure), while the remaining $n-3$ nodes are the mobile nodes (yellow nodes in the figure) that are attached to the subject/object being tracked. Within this study, UWB nodes are expected to determine inter-node distances based on Time of Flight (ToF) measurements, a commonly employed technique in commercial UWB systems [40]. Here, the ToF calculations are obtained using the Alternative Double-Sided Two-Way Ranging (AltDS-TWR) method [41], which improves accuracy by mitigating clock drift errors between devices.

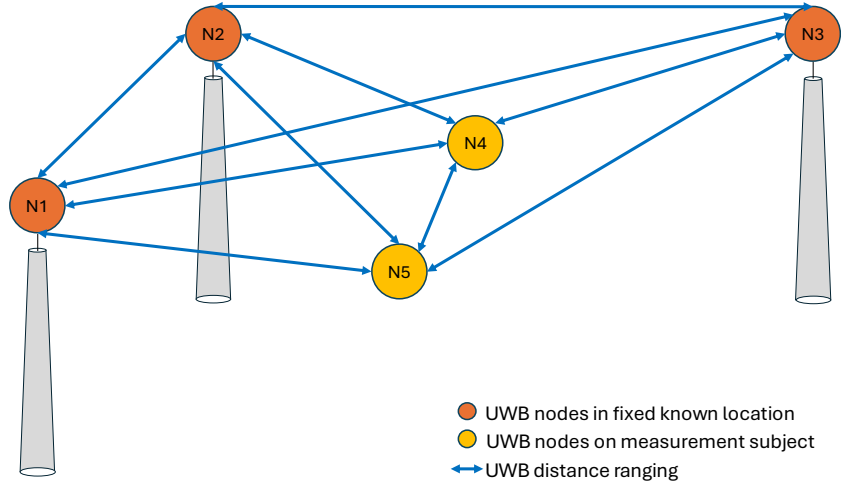

*Figure 1: UWB measurement configuration and illustration of the swarm topology.*

The UWB nodes operate in a swarm topology, where each node measures its distance to all other nodes in the system, forming a fully connected network (Figure 1). Given a total of $n$ UWB nodes in the system, where $n = n_{anchor} + n_{tag}$, with $n_{anchor}$ representing the number of fixed anchors and $n_{tag}$ representing the number of mobile tags, the number of distance measurements differs based on the topology used. For a swarm topology configuration with $n$ UWB nodes, the total number of distance measurements $M_{swarm}$ generated for each measurement cycle correspond to the triangular number of pairwise rangings and is given by:

$$M_{swarm} = \frac{n(n-1)}{2} \qquad 1$$

In comparison, the star topology configuration with the same $n$ UWB nodes generates a total number of distance estimates $M_{star}$ that is a product of the number of anchors and tags, and is given by:

$$M_{star} = n_{anchor} \times n_{tag} \qquad 2$$

For example, if there is a UWB system with 3 anchors and 2 mobile tags (a total of 5 UWB nodes), the swarm topology will provide a total of 10 unique pairwise distance measurements, while the star topology will provide a total of 6 (2 times 3) pairwise distance measurements. Therefore, the swarm topology configuration utilized in this research provides more distance estimates, which are valuable for optimization-based relative position estimation.

## 2.2 Proposed swarm position estimation

An optimization-based estimation method was implemented in this study to determine the unknown positions of $n_{tag}$ mobile sensor nodes in a swarm, based on all available inter-node UWB distance measurements, with $n_{anchor}$ fixed anchors in a network of $n$ UWB nodes ($n = n_{anchor} + n_{tag}$). The UWB distance data is initially filtered to eliminate noise, which is followed by a position estimation using a nonlinear constrained optimization approach (Figure 2). Additionally, an iterative refinement loop is included, which repeats the optimization until the RMSE of the measured distance and estimated distances from positions estimated via optimization is below a set threshold or has reached the maximum iteration count.

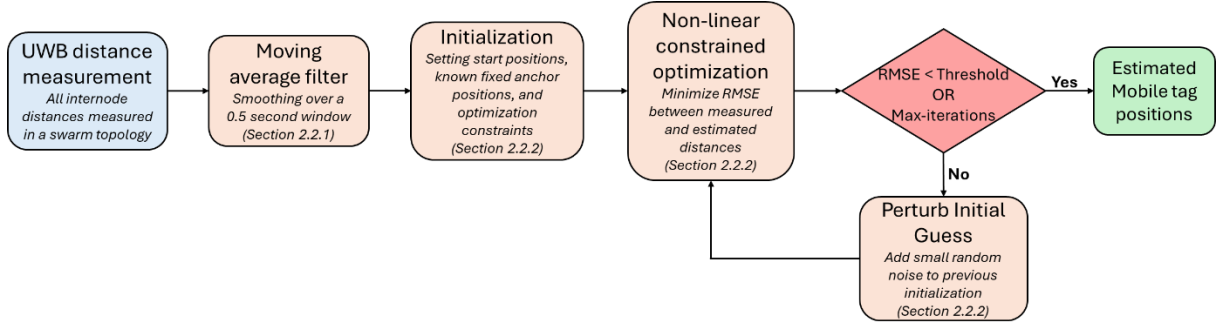

Figure 2: Proposed swarm position estimation pipeline

### 2.2.1 UWB measurement preprocessing

The UWB distance measures are initially filtered using a moving average filter to reduce the influence of high-frequency noise in UWB distance measurements. To preserve the responsiveness to actual movement, a moving average filter with a window length equivalent to 0.5 seconds was selected for smoothing. This means that for a UWB measurement system measuring at 4 samples per second, a window of 2 samples is applied, while for a UWB measurement at 20 samples per second, a window of 10 samples is applied. To eliminate phase lag, a forward and backward filtering approach was used, ensuring zero-phase distortion and accurate temporal alignment. This is equivalent to a low-pass 2nd-order filter with a Cutoff frequency of around 0.9 Hz, with zero delay or phase shift. The choice of a window of 0.5 seconds duration balances noise suppression and real-time performance, considering typical human movement velocities between 0.5 m/s and 2 m/s.

### 2.2.2 Swarm optimization-based position estimation

The position estimation of mobile UWB nodes from the distance measurements was implemented through a Sequential Quadratic Programming (SQP)-based nonlinear optimization approach (Constrained Nonlinear Optimization function, LabVIEW 2021), which was set up to find the optimal solution for the mobile sensor positions given the measured distances. For this, a set of equations was derived for the relationships between positions and distances, and an optimization criterion was defined.

The distance between any two sensor nodes $i$ and $j$, can be estimated based on the standard 3D Euclidean formula, utilizing the position coordinates of the two sensor nodes $i$ and $j$, and this estimated or reconstructed distance $\hat{d}_{ij}$ is given as:

$$\hat{d}_{ij} = \sqrt{(x_i - x_j)^2 + (y_i - y_j)^2 + (z_i - z_j)^2} \qquad 4$$

where $\hat{d}_{ij}$ represents the estimated distance between sensor $i$ and $j$, while $(x_i, y_i, z_i)$ and $(x_j, y_j, z_j)$ denote their respective 3D coordinates. This yielded a set of $n(n-1)/2$ equations, for a set of $n$ UWB nodes in the network.

The optimization objective was defined as finding the minimum of the root mean square differences between the measured and estimated positions of the mobile sensors (or the root mean square error (RMSE) in estimated position):

$$\min_{\boldsymbol{x}} f(\boldsymbol{x}) = \sqrt{\frac{1}{M} \sum_{(i,j) \in P} (d_{ij} - \hat{d}_{ij})^2} \qquad 5$$

Where $P$ is the set of all unique unordered sensor node pairs $(i, j)$ and $M$ is the number of sensor node pairs in a swarm configuration.

The measured distance values between all nodes $d_{ij}$ are preprocessed (Section 2.2.1) and used as inputs for the optimization algorithm. Depending on whether $i$ and $j$ is anchor node or mobile node, their coordinates are taken either from the known anchor positions or the optimization variable $\boldsymbol{x}$. This reconstruction of distances using the Euclidean formula results in a system of $n(n-1)/2$ nonlinear equations with $3(n-3)$ unknowns (as three fixed sensor positions are known in 3D space). In this study, in a system with 9 UWB nodes, there are 36 distance measurements, leading to 36 equations and 18 unknowns. This results in an overdetermined system with more equations than the number of unknowns and is solved through optimization.

The UWB swarm topology provides $n(n-1)/2$ unique pairwise distance measurements per update cycle for a set of $n$ UWB nodes (section 2.1). As three UWB nodes were placed at fixed, known positions (in order to accommodate both star and swarm approaches), their position values were filled out in the equations as given constants. Only the positions of the remaining $n-3$ mobile sensor nodes needed to be estimated. The unknown 3D positions of these $n-3$ mobile UWB nodes were concatenated into the optimization vector:

$$\boldsymbol{x} = [x_1, y_1, z_1, x_2, y_2, z_2, \ldots, x_{n-3}, y_{n-3}, z_{n-3}]^T \quad \in \mathbb{R}^{3(n-3)} \qquad 3$$

The formulated optimization problem was solved using the Sequential Quadratic Programming (SQP) method [42], a widely used iterative technique for constrained nonlinear problems, where each step involves solving a Quadratic Programming (QP) subproblem that locally approximates the original nonlinear objective and constraint functions. At each iteration, the QP linearizes the constraints and approximates the objective using a second-order Taylor expansion. Following this, a line search is applied for merit functions to update the estimates, and the solution is iteratively refined until convergence is achieved. This SQP method was chosen due to its effectiveness in minimizing nonlinear least-squares problems, its ability to enforce physical constraints such as the UWB node positions bounded within a defined measurement region, and its high accuracy and convergence properties in moderately sized problems like 6 unknown mobile tag positions in our tested case.

The optimization method was initialized with well-defined initial start position estimates at the first update cycle, where the start position was set at approximate sensor node locations that were determined using prior knowledge of their expected placement relative to the fixed sensor nodes. For subsequent time steps, the estimated positions from the previous time step were used as the initialization for optimization. For the experiments in this study, the search space was constrained to the actual measurement region with the minimum and maximum bounds set at [−2m, 2m] in the X and Y dimensions and at [0m, 2m] in the Z dimension.

An additional iterative refinement loop was introduced to improve robustness and avoid convergence to local minima. After each optimization cycle, the estimated positions were used to compute the reconstructed distances. If the distance estimation error exceeded a predefined threshold equal to the standard deviation of the random error component of the UWB distance measurement, the optimization was re-run with adjusted initial values. These adjustments involved adding small random perturbations (±1 cm) to the previous initialization to explore alternate solutions. This process was repeated until either the estimated distances closely matched the measured distances (i.e., RMSE fell below the threshold) or a maximum of 10 iterations was reached. If the threshold was not met within the allowed iterations, the solution with the lowest RMSE was selected as the final position estimate.

## 2.3 Trilateration-based position estimation

The standard trilateration-based approach [24] based on geometric technique was implemented for UWB position estimation, considering distance estimates that correspond to the star configuration. The UWB distance measures were initially filtered as mentioned in Section 2.2.1, to have the same input quality as used for the proposed Swarm method. In this method, the positions of each of the mobile UWB nodes were determined using their measured distances with all of the fixed sensor nodes and the known locations of the three fixed UWB nodes in the network. Since each known distance represents a sphere of possible mobile sensor positions centered at the corresponding fixed sensor node, the best estimate of the position of the intersection of these spheres in a 3D space is taken as the position estimate for the mobile sensor node. This results in a set of three Euclidean distance equation (equation 4) between each mobile node and the three known fixed sensor nodes. Solving for this set of equations provides two intersection points. Due to the positioning of the fixed nodes, only one of the two possible solutions can be identified as being inside the measurement area, and this is taken as an estimate for the position of the mobile node.

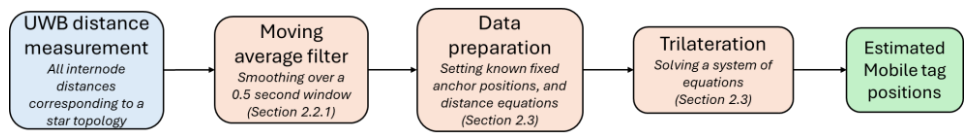

Figure 3: Trilateration-based position estimation process pipeline

## 2.4 Experimental validation

### 2.4.1 Sensitivity analysis with synthetic UWB distance estimates

The experiments were carried out in a motion capture laboratory to facilitate validation against a gold standard optical marker-based motion capture system for clinical applications (Vicon Motion Systems Ltd, Oxford, UK, 8 cameras). The optical motion capture system measures the positions of auto-reflective markers at a rate of 100 samples per second. To enable a controlled sensitivity analysis for the effect of varying levels of UWB distance measures bias and random errors, in this study the distance measurements were synthetically generated by combining the Euclidian distances between recorded position data for pairs of auto-reflective markers (simulated synthetic node positions) with synthetic bias and uniformly distributed random errors, simulating realistic distance estimates recorded through UWB nodes for different amounts of bias and random errors. Value ranges for distance estimate bias error and random error standard deviation were based on actual experiments with UWB nodes [17,43].

## 2.5 Validation Protocol

A total of nine synthetic sensor nodes were used, with three sensor nodes on fixed known locations ('anchor nodes') and the remaining six synthetic sensor nodes ('mobile nodes') were moved around. Each of the six synthetic nodes was mounted on a wooden stick, held by three persons performing varying movements. Three trials with varying free movements were performed with node distances typical for 3DAHM (0.2 to 2m), delivering position estimates of 6 mobile nodes. All measurements with the optical motion capture system were down-sampled from 100 samples per second to 4 samples per second, as this is a typical UWB update rate for a swarm of 9 nodes [15] and also delivers statistically independent samples (based on autocorrelation analysis of the position data). All the trials lasted around 5 minutes, resulting in around 1,200 position estimates per sensor per trial (12,600 position estimates for all trials). All inter-node distances for the swarm of 9 nodes were computed from the measured positions.

A sensitivity analysis was performed for position estimation performance for both the trilateration and the proposed Swarm optimization methods by comparing their estimates against the position estimates of the optical motion capture system considered as the ground-truth. This sensitivity analysis was done by repeatedly applying both methods to the synthetic node distance data with various amounts of added bias and random errors. The values for bias errors and random error standard deviation in the synthetic distance estimates were chosen based on previous experiments in which UWB node distance estimation performance was iteratively optimized and tested against the optical motion capture system [17,18]. The best performance achieved was a bias error of 0.5 cm and a random error SD of 5 cm (referred to henceforth as 'Realistic UWB distance error level'). Therefore, the sensitivity analysis was performed for this combination of error values and multiple combinations with higher and lower error values around this standard error level. Specifically, the bias in the synthetic UWB distance measurement was chosen to be discrete steps from 0 cm, 0.5 cm, 1 cm, and 2 cm. For each bias level, the standard deviation of the normally-distributed random error was progressively increased from 0 cm to 12 cm in 2 cm steps, with additional tests performed for values of 0.5 cm, 1 cm, and 5 cm. In the results section, special attention was given to the validation with the Realistic UWB distance error level.

## 2.6 Data analysis

To evaluate the accuracy of both position estimation methods, the position estimation error was calculated as the Euclidean distance between the positions estimated with these methods and those directly measured with the optical motion capture system-based positions. Mean position estimation error and its standard deviation (SD) were analyzed across various test conditions. Axis-specific accuracy was examined by analyzing the absolute mean errors and their SD in X, Y, and Z directions separately. Additionally, the statistical significance of the error reduction achieved by the swarm optimization method over trilateration in Realistic UWB distance error level conditions was tested using IBM SPSS Statistics (Version 28.0.1.0). The choice of test was determined based on the distribution of the error, where a standard paired t-test would be performed if the errors are normally distributed, or a Wilcoxon test would be performed otherwise.

# 3 Results

## 3.1 Outcome data distribution

The position estimation error based on both the proposed method and trilateration was found not to be normally distributed, therefore, a paired Wilcoxon tests were used to assess the statistical significance of the found difference in performance between both methods. The position estimation for the individual axis was normally distributed across the X and Y axes, while it was non-normal for the Z-axis. Therefore, a paired t-test was performed for the X and Y, and a Wilcoxon test for the Z-axis errors.

## 3.2 Realistic UWB distance error level validation results

The estimated positions obtained from both the proposed swarm optimization and trilateration methods for Realistic UWB distance error level (bias- 0.5 cm and random error- 5 cm) in the synthetic distance estimates from one of the trials are plotted for each coordinate axis alongside the actual ground truth positions (Figure 4). The mean 3D position estimation error for the swarm optimization method was 6.49 cm, whereas the trilateration method produced a higher mean error of 8.83 cm.

Furthermore, the SD of position estimation error was also lower for swarm optimization at 5.77 cm compared to the trilateration SD of 6.54 cm (p < 0.001). The mean difference in error between the two methods was 2.33 cm, with a 95% confidence interval ranging from 2.26 cm to 2.39 cm.

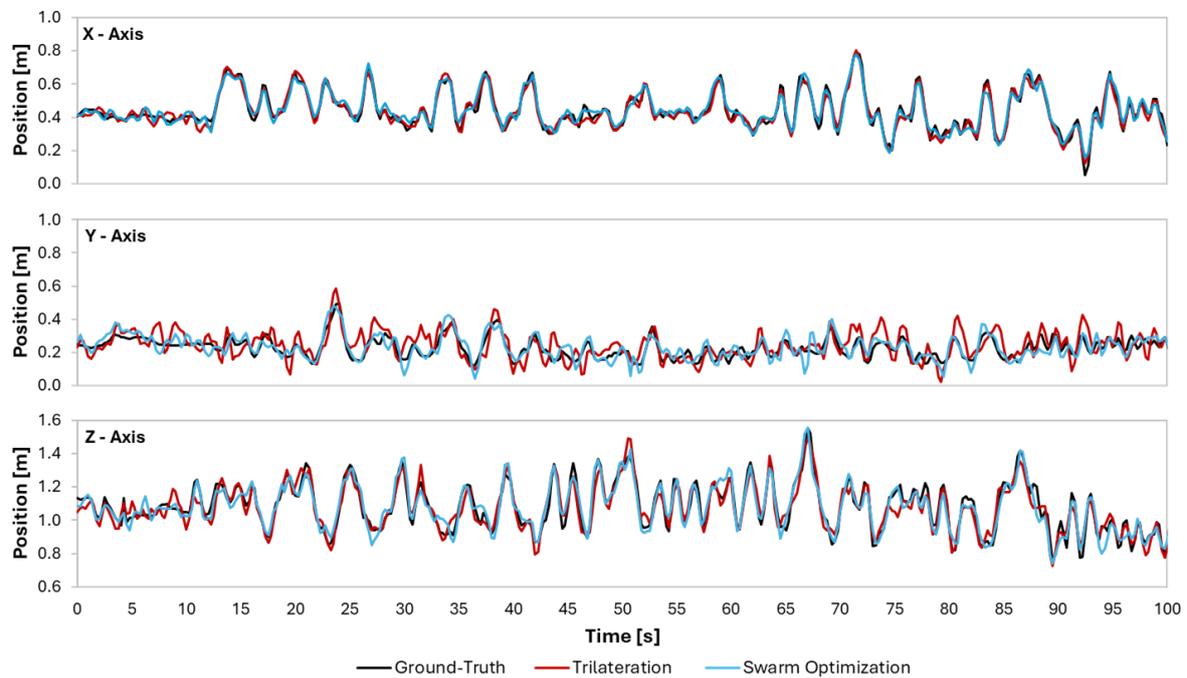

*Figure 4: A zoomed-in typical example of the position estimated by the proposed swarm optimization (blue) and trilateration (red) methods, along with the ground-truth position estimates per coordinate axis.*

Examining the errors along individual axis coordinates, the absolute mean errors along the X, Y, and Z axes for swarm optimization were 1.9 ± 3.2 cm, 3.5 ± 2.9 cm, and 4.1 ± 4.9 cm, respectively. The trilateration method exhibited higher errors, with absolute mean errors of 2.1 ± 3.3 cm, 5.9 ± 4.6 cm, and 4.7 ± 5.3 cm, respectively. The significance testing of individual axes resulted in lower p-values (p < 0.001 for x and y axes; p < 0.008 for z-axis), indicating a significant improvement over trilateration.

## 3.3   Error sensitivity analysis results

The error sensitivity analysis under a no-bias situation shows that the mean position estimation errors of the swarm optimization method rise from 0.23 cm at 0 cm random error to 13.60 cm at 12 cm random error, while the trilateration method increases more significantly from 0.23 cm to 19.55 cm over the same noise levels (Figure 5). The SD of the position estimation error follows a similar trend, with the SDs being 5.23 cm at 0 cm random error for both methods and rising to 8.38 cm for swarm optimization and 11.45 cm for trilateration at 12 cm random error. The position estimation based on the proposed swarm optimization method always showed a smaller error than the position estimates from the trilateration, except for the ideal scenario (No bias and no random error), where both had exactly the same accuracy (Figure 5). For both methods, the position estimation errors increase with the increasing random error and the bias error components in the UWB distance measurement.

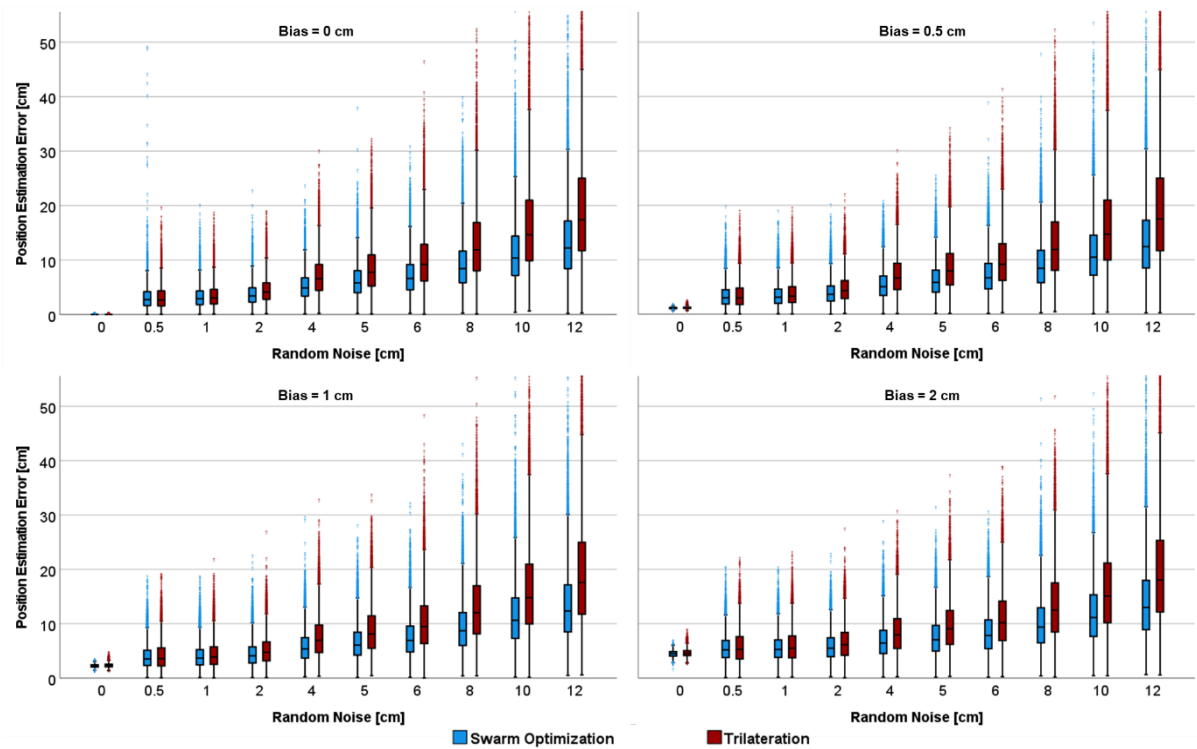

*Figure 5: Boxplots of position estimation errors for Swarm Optimization and Trilateration methods across varying levels of input random error and bias conditions. The y-axis represents the position estimation error (cm), with the graph capped at 50 cm to improve visual clarity; outliers exceeding 50 cm are not shown but were retained in the statistical analysis.*

For the swarm optimization method with no random error component in UWB distance measures, the mean position estimation errors increase from 0.23 cm at 0 cm bias to 4.66 cm at 2cm bias, while the trilateration method shows a larger increase from 0.23 cm to 4.85 cm under the same conditions. At the highest random error tested (12 cm), the error for the swarm method reaches 14.21 cm at 2 cm bias, whereas trilateration results in 19.465 cm. The significance testing indicated that for all the bias and random error cases, the improvement was always significant. Also, it can be again interpreted from the difference of the errors between the proposed swarm and trilateration method (Figure 6).

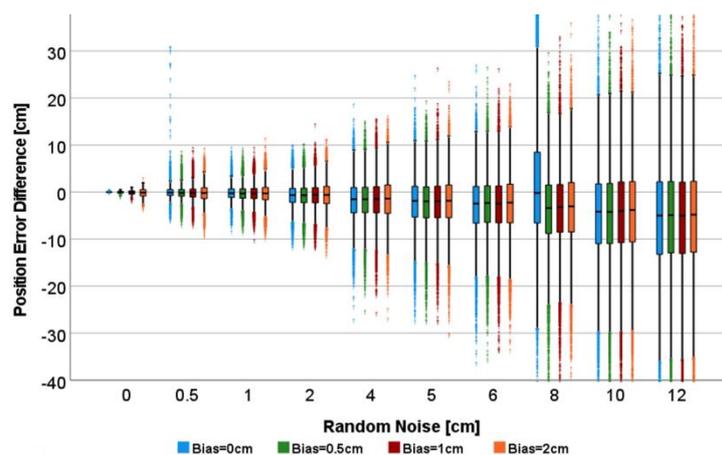

*Figure 6: Boxplot showing the distribution of the differences in position estimation errors between Swarm Optimization and Trilateration (Swarm − Trilateration) across different combinations of bias and random error levels.*

All processing time measurements were conducted on a system equipped with an Intel® Core™ i7-9700K CPU @ 3.60 GHz, with 16 GB RAM, running a 64-bit Windows 10 (Version 22H2) operating system. The positioning algorithms were implemented and executed in LabVIEW. The system was dedicated to running the positioning algorithms during the processing time measurements, with no significant background applications running. The average processing time of 21.27 ± 2.71 ms was observed for the swarm optimization method with noise combinations of no bias and low random error (None and 0.5 cm). While for all higher noise levels present in the UWB distance input, the average processing time of the swarm optimization method was 158.01 ± 57.46 ms, which was higher but very similar for all the varying noise levels. For trilateration, the average processing time per position estimate was almost the same irrespective of the input noise levels, with the mean estimation time being 0.52 ± 0.58 ms for no noise settings and 0.61 ± 0.52 ms for the highest input noise setting.

## 4 Discussion

The results of this study demonstrate that the proposed swarm optimization-based method significantly enhances position estimation accuracy compared to the conventional trilateration method. The swarm optimization-based approach method showed position estimation errors of around 6.5 cm for the Realistic UWB distance error levels in the input measurement, which is significantly lower than the trilateration method with errors of around 9 cm ($p < 0.001$). This suggests that the proposed method is a reliable alternative for UWB-based position estimation. Additionally, the SD of position estimates was always lower for the swarm-based method compared to trilateration, indicating the method has a greater consistency in its achieved accuracy.

The comprehensive error sensitivity analysis conducted in this study provides a practical understanding of the input error conditions required to achieve a desired level of position estimation accuracy. This study establishes a reference for real-world deployment of UWB-based localization methods. Error sensitivity analysis results indicate that the proposed swarm optimization method outperforms the trilateration methods in all input error conditions (Figure 5), except the zero input error scenario, where both methods performed similarly. The trilateration's performance deteriorates more rapidly as noise and bias increase, making it less reliable for higher input errors in UWB distance measures. When random error present in the UWB distance measurements is below 4 cm, trilateration performs comparably to the proposed swarm optimization method (<1 cm difference in position errors), making it a suitable solution if the required accuracy and consistency are achieved. However, for a higher noise level of 4 cm or higher, the swarm optimization method significantly outperforms trilateration by reducing errors by more than 20%, and with up to 6 cm lower error than trilateration at the highest tested noise level (i.e., around 40% reduced errors). This suggests that the proposed swarm optimization-based approach provides an overall more reliable position estimate, making it beneficial for real-world scenarios where measurement noise can be higher.

The performance demonstrated by the proposed swarm optimization-based method is comparable to a few of the more complex state-of-the-art UWB positioning algorithms [23,44,45], while better than most of the reported accuracies [29,30,32-34], tested under similar LOS conditions. For instance, [23] conducted a comparative study of multiple position estimation algorithms, where the best accuracy achieved was 5.51 cm using a Kalman filter with the least square-based position estimates utilized as an update. While all other methods tested in the study of Sang et.al., 2019, under similar conditions resulted in position estimation errors of around 8 cm [23]. Only a few cutting-edge methods push the error down to the single-digit centimeter level, specifically Guo et al., 2024 attained around 2 cm error by aggressively mitigating NLOS multipath [44], and Huang et al., 2024 achieved around 5 cm error by robust filtering (MCC-UKF fusion) in controlled settings [45]. In contrast, most real-world UWB deployments (with no sensor fusion) and based on conventional multilateration methods achieved position estimation errors in the range of 10 to 30 cm, under LOS conditions [29,30,32-34]. In essence, the proposed swarm optimization-based approach could match the best reported accuracies and exceed other conventional methods without the need for heavy infrastructure (minimal anchors) and without any complex algorithms such as sensor fusion.

In terms of computational efficiency, the trilateration method demonstrated a significantly lower computational cost, with an average processing time of less than 1 ms for each position estimate. This reflects the inherent simplicity of trilateration, which relies on direct geometric solutions without iterative computations. In contrast, the proposed swarm optimization involves solving a non-linear optimization problem with multiple variables, making it computationally more intensive by design. The proposed Swarm optimization approach exhibited an average processing time of around 158 ms per estimate. Only in the virtual absence of random error did the computational time go down substantially. Thus, while trilateration offers superior computational efficiency, the swarm optimization approach provides improved accuracy, particularly in noisy conditions, at the cost of greater algorithmic complexity and processing demands. The trade-off between accuracy and computational efficiency suggests that trilateration remains preferable in low-noise conditions, while swarm optimization is more beneficial when accuracy is a priority in noisy environments.

Across all tested conditions, the Swarm Optimization method consistently produced lower standard deviation values in the position estimation errors compared to the Trilateration method, indicating higher consistency (i.e., reduced fluctuation in the position estimates). This higher consistency of the proposed method under high-noise conditions highlights its potential for applications that require highly consistent estimates, such as clinical rehabilitation. Although the achieved accuracy is not yet sufficient for the direct use in the intended ambulatory 3D analysis of human movement and clinical rehabilitation applications, which require an accuracy around 1 cm or lower [15], the proposed method's ability to provide more consistent position estimates than the raw input UWB distances makes it a strong candidate for integration with data fusion techniques. Specifically, to achieve even lower position estimate errors, typically the UWB position estimates are used as an update in data fusion with, e.g., MIMUs. As this study shows superior UWB-based position updates with the swarm optimization-based method, it is expected that also the position estimates of such data fusion methods should show lower errors than the current state of the art.

Though the extensive error sensitivity analysis provides insights into error propagation, guiding future application of this method, however, the real-life potential of the proposed method still has to be confirmed in experiments with actual UWB data. Especially, the errors caused by NLOS conditions and their mitigation should be studied in detail. Also, further improvements should focus on optimally exploiting the redundancy in the position estimates in a fully connected swarm to balance accuracy, mitigation of NLOS-related errors, update rate, and computational load.

## 5 Conclusion

This study introduces and evaluates a swarm optimization-based position estimation method that utilizes all available internode distances rather than restricting measurements to a star topology. The proposed swarm optimization approach was validated against an optical measurement system together with a well-established trilateration method, and their performance was compared. The swarm optimization method significantly improves the position estimation over the trilateration method, with position estimation errors around 6.5 cm. An extensive error sensitivity analysis provided a deeper understanding of how the error characteristics of UWB distance estimates influence position estimation quality. It is expected that using the improved position estimates in data fusion, e.g., with MIMUs, will improve the data fusion position estimation quality. Further research is required to achieve the lower error level required for the intended high-accuracy applications, such as 3D AHM.

The increased accuracy in UWB-based position estimation seems generally beneficial for any application using a UWB swarm.

**Author Contributions**

**Vinish Yogesh:** Conceptualization, Methodology, Validation, Formal Analysis, Writing—Original draft preparation **Bert-Jan van F. Beijnum:** Supervision, Writing—Review and Editing **Jaap H. Buurke:** Supervision, Writing—Review and Editing **Chris T. M. Baten:** Conceptualization, Supervision, Writing—Review and Editing, Project Administration, Funding Acquisition

**Funding**

This research was funded by the European Fund for Regional Development (NL-EFRO) under grant number PROJ-000965.

## 6 References

1. Cheraghinia, M.; Shahid, A.; Luchie, S.; Gordebeke, G.-J.; Caytan, O.; Fontaine, J.; Herbruggen, B.V.; Lemey, S.; Poorter, E.D. A Comprehensive Overview on UWB Radar: Applications, Standards, Signal Processing Techniques, Datasets, Radio Chips, Trends and Future Research Directions. *IEEE Communications Surveys & Tutorials* **2024**, 1-1, doi:10.1109/comst.2024.3488173.
2. Hapsari, G.I.; Munadi, R.; Erfianto, B.; Irawati, I.D. Future Research and Trends in Ultra-Wideband Indoor Tag Localization. *IEEE Access* **2024**, 1-1, doi:10.1109/access.2024.3399476.
3. Zafari, F.; Gkelias, A.; Leung, K.K. A Survey of Indoor Localization Systems and Technologies. *IEEE Communications Surveys & Tutorials* **2019**, *21*, 2568-2599, doi:10.1109/comst.2019.2911558.
4. Jie, H.; Yishuang, G.; Pahlavan, K. Toward Accurate Human Tracking: Modeling Time-of-Arrival for Wireless Wearable Sensors in Multipath Environment. *IEEE Sensors Journal* **2014**, *14*, 3996-4006, doi:10.1109/jsen.2014.2356857.
5. Lyu, Y.; Wei, M.; Li, S.; Wang, D. A fusion positioning system with environmental-adaptive algorithm: IPSO-IAUKF fusion of UWB and IMU for NLOS noise mitigation. *Measurement: Sensors* **2025**, *38*, doi:10.1016/j.measen.2025.101864.
6. Haggenmiller, A.; Krogius, M.; Olson, E. Non-parametric Error Modeling for Ultra-wideband Localization Networks. In Proceedings of the 2019 International Conference on Robotics and Automation (ICRA), Montreal, Canada, 20/05/2019, 2019; pp. 2568-2574.
7. Cheng, L.; Fu, Z. An adaptive Kalman filter loosely coupled indoor fusion positioning system based on inertial navigation system and ultra-wide band. *Measurement* **2025**, *244*, doi:10.1016/j.measurement.2024.116412.
8. Wang, X.; Gao, F.; Huang, J.; Xue, Y. UWB/LiDAR Tightly Coupled Positioning Algorithm Based on ISSA Optimized Particle Filter. *IEEE Sensors Journal* **2024**, *24*, 11217-11228, doi:10.1109/jsen.2024.3366941.
9. Xu, Y.; Chen, Z.; Zhao, M.; Tang, F.; Li, Y.; Liu, J.; Kato, N. UVtrack: Multi-Modal Indoor Seamless Localization Using Ultra-Wideband Communication and Vision Sensors. *IEEE Open Journal of the Computer Society* **2025**, *6*, 272-281, doi:10.1109/ojcs.2025.3531442.
10. Hashim, H.A.; Eltoukhy, A.E.E.; Vamvoudakis, K.G. UWB Ranging and IMU Data Fusion: Overview and Nonlinear Stochastic Filter for Inertial Navigation. *IEEE Transactions on Intelligent Transportation Systems* **2024**, *25*, 359-369, doi:10.1109/tits.2023.3309288.
11. Feng, D.; Wang, C.; He, C.; Zhuang, Y.; Xia, X.G. Kalman-Filter-Based Integration of IMU and UWB for High-Accuracy Indoor Positioning and Navigation. *IEEE Internet of Things Journal* **2020**, *7*, 3133-3146, doi:10.1109/JIOT.2020.2965115.
12. Chow, J.; Hol, J.; Luinge, H. Tightly-Coupled Joint User Self-Calibration of Accelerometers, Gyroscopes, and Magnetometers. *Drones* **2018**, *2*, doi:10.3390/drones2010006.

13. Kaichi, T.; Maruyama, T.; Tada, M.; Saito, H. Resolving Position Ambiguity of IMU-Based Human Pose with a Single RGB Camera. *Sensors (Basel)* **2020**, *20*, doi:10.3390/s20195453.
14. Li, X.; Wang, Y.; Liu, D. Research on Extended Kalman Filter and Particle Filter Combinational Algorithm in UWB and Foot-Mounted IMU Fusion Positioning. *Mobile Information Systems* **2018**, *2018*, doi:10.1155/2018/1587253.
15. Yogesh, V.; Buurke, J.H.; Veltink, P.H.; Baten, C.T.M. Integrated UWB/MIMU Sensor System for Position Estimation towards an Accurate Analysis of Human Movement: A Technical Review. *Sensors* **2023**, *23*, doi:10.3390/s23167277.
16. Al-Okby, M.F.R.; Junginger, S.; Roddelkopf, T.; Thurow, K. UWB-Based Real-Time Indoor Positioning Systems: A Comprehensive Review. *Applied Sciences* **2024**, *14*, doi:10.3390/app142311005.
17. Yogesh, V.; Grevinga, L.; Voort, C.; Buurke, J.H.; Veltink, P.H.; Baten, C.T.M. Novel calibration method for improved UWB sensor distance measurement in the context of application for 3D analysis of human movement. *Engineering Science and Technology, an International Journal* **2024**, *58*, doi:10.1016/j.jestch.2024.101844.
18. APS014 Application Note, Antenna Delay Calibration of DW1000-Based Products and Systems. Available online: https://www.qorvo.com/products/d/da008449 (accessed on June 25, 2023).
19. Shah, S.; Chaiwong, K.; Kovavisaruch, L.-O.; Kaemarungsi, K.; Demeechai, T. Antenna Delay Calibration of UWB Nodes. *IEEE Access* **2021**, *9*, 63294-63305, doi:10.1109/access.2021.3075448.
20. Sidorenko, J.; Schatz, V.; Scherer-Negenborn, N.; Arens, M.; Hugentobler, U. Decawave UWB Clock Drift Correction and Power Self-Calibration. *Sensors (Basel)* **2019**, *19*, doi:10.3390/s19132942.
21. Cano, J.; Pages, G.; Chaumette, E.; LeNy, J. Clock and Power-Induced Bias Correction for UWB Time-of-Flight Measurements. *IEEE Robotics and Automation Letters* **2022**, *7*, 2431-2438, doi:10.1109/lra.2022.3143202.
22. Shalaby, M.A.; Cossette, C.C.; Forbes, J.R.; Ny, J.L. Calibration and Uncertainty Characterization for Ultra-Wideband Two-Way-Ranging Measurements. In Proceedings of the 2023 IEEE International Conference on Robotics and Automation (ICRA), London, United Kingdom, 2023; pp. 4128-4134.
23. Sang, C.L.; Adams, M.; Hesse, M.; Hörmann, T.; Korthals, T.; Rückert, U. A comparative study of UWB-based true-range positioning algorithms using experimental data. In Proceedings of the 2019 16th Workshop on Positioning, Navigation and Communications (WPNC), 2019; pp. 1-6.
24. Sahinoglu, Z.; Gezici, S.; Güvenc, I. *Ultra-wideband positioning systems: theoretical limits, ranging algorithms, and protocols*; Cambridge university press: 2008.
25. Jing, L.; Zhang, J.-F. LS-Based Parameter Estimation of DARMA Systems with Uniformly Quantized Observations. *Journal of Systems Science and Complexity* **2021**, *35*, 748-765, doi:10.1007/s11424-021-0314-y.
26. Shu, B.; Li, C.; Wang, H.; Li, H. An Improved Chan-Taylor Hybrid Location Algorithm Based on UWB. In Proceedings of the 2024 6th International Conference on Electronic Engineering and Informatics (EEI), 2024; pp. 1783-1789.
27. Han, H.; Wang, J.; Liu, F.; Zhang, J.; Yang, D.; Li, B. An Emergency Seamless Positioning Technique Based on ad hoc UWB Networking Using Robust EKF. *Sensors (Basel)* **2019**, *19*, doi:10.3390/s19143135.
28. Liu, A.; Wang, J.; Lin, S.; Kong, X. A Dynamic UKF-Based UWB/Wheel Odometry Tightly Coupled Approach for Indoor Positioning. *Electronics* **2024**, *13*, doi:10.3390/electronics13081518.
29. Fontaine, J.; Che, F.; Shahid, A.; Van Herbruggen, B.; Ahmed, Q.Z.; Bin Abbas, W.; De Poorter, E. Transfer Learning for UWB Error Correction and (N)LOS Classification in Multiple

30. Waqar, A.; Ahmad, I.; Habibi, D.; Phung, Q.V. Analysis of GPS and UWB positioning system for athlete tracking. *Measurement: Sensors* **2021**, *14*, doi:10.1016/j.measen.2020.100036.
31. Guo, H.; Li, M.; Zhang, X.; Gao, X.; Liu, Q. UWB indoor positioning optimization algorithm based on genetic annealing and clustering analysis. *Front Neurorobot* **2022**, *16*, 715440, doi:10.3389/fnbot.2022.715440.
32. Cretu-Sircu, A.L.; Schioler, H.; Cederholm, J.P.; Sircu, I.; Schjorring, A.; Larrad, I.R.; Berardinelli, G.; Madsen, O. Evaluation and Comparison of Ultrasonic and UWB Technology for Indoor Localization in an Industrial Environment. *Sensors (Basel)* **2022**, *22*, doi:10.3390/s22082927.
33. Cho, J.; Jeong, S.; Lee, B. A Study on Anchor Placement and 3D Positioning Algorithm for UWB Application in Small Sites. *KSCE Journal of Civil Engineering* **2024**, *28*, 4575-4587, doi:10.1007/s12205-024-2107-z.
34. Zhao, M.; Chang, T.; Arun, A.; Ayyalasomayajula, R.; Zhang, C.; Bharadia, D. ULoc. *Proceedings of the ACM on Interactive, Mobile, Wearable and Ubiquitous Technologies* **2021**, *5*, 1-31, doi:10.1145/3478124.
35. Zhou, N.; Si, M.; Li, D.; Seow, C.K.; Mi, J. An Indoor UWB 3D Positioning Method for Coplanar Base Stations. *Sensors (Basel)* **2022**, *22*, doi:10.3390/s22249634.
36. Jimenez, A.R.; Seco, F. Improving the Accuracy of Decawave's UWB MDEK1001 Location System by Gaining Access to Multiple Ranges. *Sensors (Basel)* **2021**, *21*, doi:10.3390/s21051787.
37. Volpi, A.; Tebaldi, L.; Matrella, G.; Montanari, R.; Bottani, E. Low-Cost UWB Based Real-Time Locating System: Development, Lab Test, Industrial Implementation and Economic Assessment. *Sensors (Basel)* **2023**, *23*, doi:10.3390/s23031124.
38. Li, J.; Xue, J.; Fu, D.; Gui, C.; Wang, X. Position Estimation and Error Correction of Mobile Robots Based on UWB and Multisensors. *Journal of Sensors* **2022**, *2022*, doi:doi:10.1155/2022/7071466.
39. Waqar, A.; Ahmad, I.; Habibi, D.; Phung, Q.V. A range error reduction technique for positioning applications in sports. *The Journal of Engineering* **2021**, *2021*, 73-84, doi:10.1049/tje2.12010.
40. Alavi, B.; Pahlavan, K. Modeling of the TOA-based distance measurement error using UWB indoor radio measurements. *IEEE communications letters* **2006**, *10*, 275-277, doi:10.1109/LCOMM.2006.1613745.
41. Neirynck, D.; Luk, E.; McLaughlin, M. An Alternative Double-Sided Two-Way Ranging Method. In Proceedings of the 2016 13th workshop on positioning, navigation and communications (WPNC), 2016; pp. 1-4.
42. Nocedal, J.; Wright, S.J. Numerical optimization. **2006**.
43. Yogesh, V.; Rook, J.W.A.; Keizers, T.; Voort, C.; Buurke, J.H.; Veltink, P.H.; Baten, C.T.M. UWB distance estimation errors in (non-)line of sight situations within the context of 3D analysis of human movement. *Engineering Research Express* **2024**, *6*, doi:10.1088/2631-8695/ad7e7e.
44. Guo, H.; Song, S.; Yin, H.; Ren, D.; Zhu, X. Optimization of UWB indoor positioning based on hardware accelerated Fuzzy ISODATA. *Sci Rep* **2024**, *14*, 17985, doi:10.1038/s41598-024-68998-0.
45. Huang, Y.; Cao, B.; Wang, A. Design a novel algorithm for enhancing UWB positioning accuracy in GPS denied environments. *Sci Rep* **2024**, *14*, 23895, doi:10.1038/s41598-024-74773-y.